\documentclass[doublecol]{epl2}
\usepackage{amsmath}
\usepackage{amssymb,amsthm}
\usepackage{graphicx}

\title{Photon and Photon-Added Intelligent States of Coupled Parametric Oscillators}

\author{A.E. Neira \and R. Mulet}

\institute{                    
Group of Complex Systems and Statistical Physics, Department of Theoretical Physics, University of Havana. La Habana, Cuba.}
\pacs{nn.mm.xx}{First pacs description}
\pacs{nn.mm.xx}{Second pacs description}
\pacs{nn.mm.xx}{Third pacs description}

\abstract{
  We study a quantum system of coupled oscillators subject to a periodic excitation
  of its parameters. Using Floquet-Lyapunov theory we derive the linear
  integrals of motion of the system and relate their covariance matrix 
  to that for the canonical observables. The operator 
  integrals allows us to construct the intelligent (minimum uncertainty) states
  of the system and the corresponding photon-added states. We found explicit 
  expressions for the wavefunction, Wigner function and covariance matrix of these states.
}

\begin{document}

\maketitle

\section{Introduction}

The understanding of the quantum harmonic oscillator is one of the basic ideas to be mastered by any Physics graduate student. Probably every text book
in Quantum Mechanics devotes at least one full chapter to introduce and discuss the model. Therefore, it may appear surprising that the introduction of time-dependent parameters turns the model into an active field of research \cite{ncoupldho,tdhopointtrans,landa,wolf1,tdhocs,holz,manko}.

Indeed, the study of quantum harmonic oscillators with explicitly time-dependent parameters\cite{wolf1} has been stimulated by their effectiveness in the description of a wide range of quantum phenomena,  like propagation of light through variable media \cite{variablemedia}, the dynamical Casimir effect \cite{kashmir} and the motion of trapped ions \cite{rmpblatt}. The latter, also a popular proposal for scalable quantum computation \cite{quantumcomputation}.

Broadly speaking, the explicit time-dependence of the parameters in the model induces non-vanishing covariances between the canonical operators and changes the distribution of excitations of the states of the system \cite{manko}. Moreover, when the coupling between two non-stationary oscillators \cite{kalmykov} is considered, the inseparability of the system's wave-functions takes a stronger character. In general, a coordinate transformation that gives real uncoupled modes of oscillation is no longer possible and quantum entanglement may show up \cite{lizuain}. This entanglement turns out to be a property of the algebraic structure of the model and not just of the states.

In this work we study a system  of $N$ parametric oscillators described by the Hamiltonian
\begin{equation}\label{hamiltonian}
	\hat{H}(t)=\frac{1}{2}\hat{\textbf{x}}^TJ\Pi(t)\hat{\textbf{x}}
\end{equation}
where $\hat{\textbf{x}}$ is an $2N$ dimensional vector whose first $N$ components correspond to the position and the last $N$ to the momentum operator of each oscillator. The parametric excitation is described by the explicit time-dependence of the frequencies and coupling factors present in $\Pi(t)$. In particular, we will focus our attention in a class of models in which $\Pi(t)=\Pi(t+T)$. The canonical commutation relations for the observables, written in matrix form and using natural units $\hbar\equiv1$, read
\begin{equation}
	\left[\hat{x}_i,\hat{x}_j\right]=iJ_{ij}.
\end{equation}
The two $2N\times2N$ matrices appearing on the right hand side of \eqref{hamiltonian} are, in general, given by
\begin{equation}
	\begin{array}{ccc}
		J=\left(\begin{array}{cc}
			0 & 1_N \\
			-1_N & 0
		\end{array}
		\right),& &\Pi(t)=\left(\begin{array}{cc}
			C_{q\leftrightarrow p} & C_{p\leftrightarrow p} \\
			C_{q\leftrightarrow q} & C_{q\leftrightarrow p}
		\end{array}
		\right),
	\end{array}
\end{equation}
which means that $J$ is a skew-symmetric (symplectic) matrix $J^T=J^{-1}=-J$ and $\Pi(t)$ describes the configuration of the system. That is, its $C_{q\leftrightarrow q}$ block contains the frequencies and coupling factors between the coordinates, $C_{p\leftrightarrow p}$ represents the coupling between momenta and $C_{q\leftrightarrow p}$ the coupling between positions and momenta. 

A major concern in the study of these parametrically excited oscillators is the obtention of their coherent or classical states. Such classicality is determined by a positive Wigner function and by the Poissonian character of the photon distribution. The canonical coherent states \cite{glauber} of a stationary harmonic oscillator are known to minimize the Heisenberg uncertainty relation \cite{heisenberg,kennard}. But the family of coherent states is larger, they can be constructed not only on the simple one-dimensional case, but for coupled oscillators \cite{kalmykov,kalmykov2} or further generalizations of the quadratic Hamiltonian model. For example, the parametric excitation of a single harmonic oscillator induces squeezing of the quadrature components and correlations between them. In this case, the coherent states become squeezed and correlated \cite{manko,kalmykov2}, and minimize Schr\"odinger's uncertainty relation \cite{schrodinger}. Such minimization of uncertainty inequalities \cite{heisenberg,kennard,schrodinger,robertson} is the defining property of the intelligent states of a quantum mechanical system, of which the coherent states are a paramount example.

The intelligent states, in general, belong to the Gaussian regime, that is, their Wigner function is a Gaussian distribution and occupy a privileged position in quantum information. From the experimental side, these are the states that can be most easily obtained in the laboratory and this has also called the attention of many theoreticians. However, the departure from the Gaussian regime might provide a better playground for quantum information processing protocols with no classical analog \cite{continousvariableqc} and motivates the search for new analytic methods and solutions.


The non-Gaussian regime imposes non classical behavior. Non-classical states are generally defined as the ones with negative values of the Wigner function and sub- or super-poissonian photon distribution function \cite{mandelparameter}. Within these classes we can identify the excited or photon-added coherent states. Their behavior is intermediate between that of classical and more general quantum states and provide a scheme to study the transition from particle-like to wave-like behavior of light \cite{experimentalPACS}. These states were firstly introduced by Agarwal and Tara \cite{agarwal} and have gained popularity in the last three decades due to their possible implementations in quantum information processing. In the context of oscillators systems, they have been previously studied for a single oscillator with a generic time-dependence of its parameters \cite{dodonovPacs} and for a generic Gaussian state in the case of a single-photon addition \cite{MATTIA}. 

Our goal in this work is to find the photon and the photon-added intelligent states for a system of parametrically coupled periodic oscillators. The remainder of the manuscript is organized in the following way. We first construct the linear integrals of motion for \eqref{hamiltonian} exploiting well known results from Floquet theory.  We will show that these operators translate into creation and annihilation operators and they will be used to introduce the non-stationary Fock basis spanning the Hilbert space of the oscillators. Then, we show that these Fock states do not minimize, in general, the Robertson uncertainty relation. We continue introducing a class of intelligent states by performing a unitary transformation on the ground state, which will be in general squeezed and correlated. Then, adding excitations to the previously introduced intelligent states, we obtain the wavefunction, Wigner function and the corresponding covariance matrix of the excited states. Finally, we present the conclusions of our work.

\section{Floquet Theory and Linear Integrals of Motion}
In this section we make use of the periodicity in time of the Hamiltonian operator \eqref{hamiltonian} to employ the Floquet theorem in the construction of the linear integrals of motion of the quantum mechanical problem. For completeness we will review the definition of the time-dependent transformation (Floquet-Lyapunov) and the Heisenberg's creation and annihilation operators following reference \cite{landa}.

Let us start transforming the Hamiltonian \eqref{hamiltonian} to its Heisenberg picture equivalent
\begin{equation}\label{heisenberghamiltonian}
	\hat{H}_{h}(t)=\frac{1}{2}\hat{\textbf{x}}^T(t)J\Pi(t)\hat{\textbf{x}}(t),
\end{equation}
where the $h$ subscript stands for Heisenberg and the components of $\hat{\textbf{x}}(t)$ are Heisenberg operators constructed from the $\hat{\textbf{x}}$ of the Schr\"odinger picture. The equations of motion derived from \eqref{heisenberghamiltonian} take the form
\begin{equation}\label{heisenbergeom}
	\dot{\hat{x}}_k(t)=-i\left[\hat{x}_k(t),\hat{H}_{Hei}(t)\right]=\sum_{l}\Pi_{kl}(t)\hat{x}_l(t)
\end{equation}
This is \emph{formally} identical to a classical linear dynamical system with periodic coefficients. Hence, according to Floquet theory \cite{gfloquet}, they may be described in terms of independent complex modes of oscillation $\hat{\chi}(t)$ (Floquet modes), related to $\hat{\textbf{x}}(t)$ through the Floquet-Lyapunov transformation (FLT), $F(t)$: 
\begin{equation}\label{FLT}
	\hat{\textbf{x}}_{h}(t)=F(t)\hat{\chi}(t).
\end{equation}

The time evolution of the operators $\hat{\chi}(t)$ is entirely described, in the Heisenberg picture, by imaginary exponential functions
\begin{equation*}
	\hat{\chi}_j(t)=\hat{\chi}_j(0)\exp(i \omega_j t) 
\end{equation*} 
\begin{equation}\label{floquetmodes}
	\hat{\chi}_{j+N}(t)=\hat{\chi}_{j+N}(0)\exp(-i \omega_{j} t)
\end{equation}
where the $j$ index runs from $1$ to $N$ and the $\omega_j$ are the Floquet exponents of the system. These are the quantized Floquet modes of the classical system. The time-dependent transformation $F(t)$ shares the period of the Hamiltonian \eqref{hamiltonian}, and its elements satisfy
\begin{equation}\label{ffconjugate}
	F_{i,j+N}(t)=F_{i,j}^*(t),\hspace*{0.3cm}j=1,...,N\hspace*{0.1cm}i=1,...,2N
\end{equation}
Using this last expression we can rewrite the Floquet-Lyapunov transformation in block form
\begin{equation}\label{FLT2}
	F(t)=\left(\begin{array}{cc}
		U & U^* \\
		V & V^*
	\end{array}
	\right).
\end{equation}
These blocks are time-dependent $N\times N$ matrices which share the period of the parameters. They constitute a generalization of the periodic functions appearing in the Floquet theorem \cite{maclahan,gfloquet} to the case of several degrees of freedom.

The time-dependent transformation \eqref{FLT} implicitly depends on the configuration $\Pi(t)$ of the system. In the simple case of uncoupled oscillators, every block of $\Pi(t)$ is a diagonal matrix, this implies that the blocks of $F(t)$, that is, $U$, $V$ and their complex conjugates, are also diagonal, $F_{ij}=0$ if $i\neq j$. The coupling between the canonical variables would then be reflected in non-vanishing off-diagonal elements of the blocks. We assume non zero coupling throughout this work.

If we impose the following condition to the FLT,
\begin{equation}\label{cannonicalcondition}
 	\sum_{l=1}^N F_{l+N,k}(0)F_{l,m}(0)=\frac{i}{2}\delta_{lm},
\end{equation}
or, equivalently, $V^T(0)U(0)=\frac{i}{2}I$, then the transformation will be canonical \cite{landa,wolf2}. 
The above condition will also ensure that the quantized Floquet modes (operators in the Heisenberg picture), satisfy the Weyl-Heisenberg algebra. Written in matrix form, this is
\begin{equation}\label{weylhei}
	\left[\hat{\chi}_{i}(t),\hat{\chi}_{j}(t)\right]=-J_{ij},
\end{equation} 
and, taking \eqref{floquetmodes} into account, we can identified the Floquet modes, in the Heisenberg picture, with creation ($\hat{\chi}_i(t)$) and annihilation ($\hat{\chi}_{i+N}(t)$, $i=1,...,N$) operators. We will assume the fulfillment of \eqref{cannonicalcondition} from now on.
  

In the Schr\"odinger picture,
\begin{equation}\label{FLTschrödinger}
	\hat{\textbf{x}}=F(t)\hat{\chi}_S(t).
\end{equation}
But the unitarity of the time evolution operator ensures the invariance of the Weyl-Heisenberg algebra under its action. This, in turns guarantees that also  the operators obtained from the Floquet modes in the Schr\"odinger picture satisfy the commutation relations \eqref{weylhei}.  Then, from these Schr\"odinger operators we can define the following (explicit) time-dependent operators
\begin{equation}\label{IOM}
	\left(\begin{array}{c}
		\hat{A}^\dagger(t)\\
		\hat{A}(t)
	\end{array}
	\right)=\left(\begin{array}{cc}
		e^{-i\Omega t} & 0 \\
		0 & e^{i\Omega t}
	\end{array}
	\right)\hat{\chi}_S(t)\equiv\hat{I}(t)
\end{equation}
where $\Omega=diag(\omega_1,...,\omega_N)$. 
Substituting \eqref{IOM} in \eqref{FLTschrödinger} and using the inverse of the FLT (see \cite{landa}) we can express $\hat{A}(t)$ and $\hat{A}^\dagger(t)$ in terms of the time-independent position and momentum operators
\begin{equation}
	\hat{A}^{\dagger}(t)=iV^{\dagger}(t)e^{-i\Omega t}\hat{q}-iU^{\dagger}(t)e^{-i\Omega t}\hat{p},
\end{equation}
\begin{equation}\label{liomintermsofqp}
	\hat{A}(t)=-iV^T(t)e^{i\Omega t}\hat{q}+iU^T(t)e^{i\Omega t}\hat{p}.
\end{equation}
It is easy to see that the total derivative of this last expression vanishes, and the following operator equation is satisfied
\begin{equation}\label{integralsofmotion}
	\dfrac{\partial\hat{I}_k(t)}{\partial t}=i\left[\hat{I}_k(t),\hat{H}(t)\right],
\end{equation}
where $\hat{H}(t)$ is the Schr\"odinger Hamiltonian \eqref{hamiltonian}. Hence, the time-dependent non-Hermitian operators $\hat{A}^\dagger(t)$ and $\hat{A}(t)$ (group together in $\hat{I}(t)$), are integrals of motion for the quantum mechanical problem of the Dodonov-Malkin-Man'ko type \cite{dodonovLIOM,manko} and can be used  to define an orthonormal basis for the Hilbert space of the oscillators. Moreover, since they satisfy the boson commutation relations \eqref{weylhei},  we can reinterpret each oscillator as a bosonic system in the second quantization formalism.

The construction of an orthonormal basis is straightforward: We assume that for $t\leq0$ the system is a stationary one, in such a situation the configuration matrix can be redefine as $\Pi(t)\rightarrow(\Pi(0)\Theta(-t)+\Pi(t)\Theta(t))$, where $\Theta(t)$ is the Heaviside step function. Then, for $t\leq0$, the integrals of motion are just standard creation and annihilation operators and the Fock basis $|{n}\rangle$ can be constructed in the usual way. For $t\geq 0$ the action of the time evolution operator $\hat{\mathcal{U}}(t)=\exp(\frac{i}{\hbar}\int_{0}^{t}H(t')dt')$ upon the stationary Fock basis defines the \emph{dynamical Fock states} \cite{rmpblatt} of the parametrically excited system,
\begin{equation}\label{dynfock}
	|\{n\};t\rangle=\hat{\mathcal{U}}(t)|\{n\}\rangle
\end{equation}
\noindent where $\{n\}$ is the set $\{n_1,...,n_N\}$. 
We define, in the Schr\"odinger picture, the number operator for the $i-$th oscillator ($i<N$) as $\hat{n}_i(t)=\hat{A}_i^\dagger(t)\hat{A}_i(t)$.
Its eigenvectors are the states defined in \eqref{dynfock} and the explicit eigenvalue problem takes the form
\begin{equation}
	\hat{n}_i(t)|\{n\};t\rangle=n_i|\{n\};t\rangle,
\end{equation}
This implies that \eqref{dynfock} describes the states that evolved from a stationary state $|\{n\}\rangle$ with $n_i$, $i=1,...,N$ excitations in the i-th mode. In a quantum optical setting nonzero coupling will be translated in entangled modes, and the dynamical Fock states, as well as all those to be introduced in this letter will be multimode states. For definiteness we will be referring to the excitations as photons even if the oscillators system is not necessarily restricted to the modes of the radiation field.

The action of the linear integrals of motion on \eqref{dynfock} is analogous to the action of creation and annihilation operators on the Fock basis for the stationary problem. The explicit expressions read
\begin{equation}\label{aonfock}
	\hat{A}_i(t)|\{n\};t\rangle=\sqrt{n_i}|...n_i-1...;t\rangle,
\end{equation}
\begin{equation}\label{adagaonfock}
	\hat{A}_{i}^\dagger(t)|\{n\};t\rangle=\sqrt{n_i+1}|...n_i+1...;t\rangle.
\end{equation}

We would like to emphasize that, for the previous treatment to hold it is necessary that the Floquet exponents $\omega_j$ of the system are strictly real numbers. This is equivalent to say that the elements of $\Pi(t)$ have to produce stable oscillations for the classical system.

\section{Robertson Inequality and Intelligent States}

The mathematical description of the uncertainty principle for a set of $2 N$ observables of a certain quantum system can be given through the Robertson inequality \cite{robertson},
\begin{equation}\label{intelligencecondition}
	\textbf{Det}(\sigma(\hat{x}))\geq\left(\frac{1}{2}\right)^{2N}.
\end{equation}
In the left-hand side figures the determinant of covariance (dispersion) matrix $\sigma$ corresponding to the systems's quantum mechanical state. We will take the set to be the $2N$ components of the vector $\hat{\textbf{x}}$. Then, the elements of its covariance matrix, in a state $|\psi;t\rangle$, are given by
\begin{equation}\label{covarianceelements}
	\sigma(\hat{x}_i,\hat{x}_j)=\frac{1}{2}\langle\left\{\hat{x}_i,\hat{x}_j\right\}\rangle_{\psi;t}-\langle\hat{x}_i\rangle\langle\hat{x}_j\rangle_{\psi;t}
\end{equation}
where $\left\{\cdot,\cdot\right\}$ denotes the anti-commutator. Identical expressions hold for the integrals of motion $\hat{I}_{i}(t)$. 
In fact, the covariance matrix of the canonical observables $\sigma(\hat{x}(t))$ can be related to that of the integrals of motion $\sigma(\hat{I}(t))$ through the Floquet-Lyapunov transformation
\begin{equation}\label{sigmaF}
	\sigma(\hat{\textbf{x}})=F(t)\exp\left(iWt\right)\sigma(\hat{I}(t))\exp\left(iWt\right)F^T(t),
\end{equation}
where 
\begin{equation*}
	W=\left(\begin{array}{cc}
		\Omega & 0 \\
		0 & -\Omega
	\end{array}
	\right)
\end{equation*}
Expression \eqref{sigmaF} follows directly from \eqref{FLTschrödinger} and \eqref{IOM}. Note that, although $\hat{I}(t)$ are time-dependent, $\sigma(\hat{\chi}(t))$ is not necessarily time-dependent too. On the other hand, $\sigma(\hat{x})$ depends explicitly on time, due to the presence of the Floquet-Lyapunov transformation and $\exp(iWt)$ in \eqref{sigmaF}. 

The columns of the matrix $F\exp(iWt)$ correspond to the solutions of the classical system of coupled parametric oscillators subject to the initial conditions \eqref{cannonicalcondition}. This means that the covariances of canonical observables, in any quantum mechanical state, are determined by the covariances of $\hat{I}(t)$ and by the classical motion of the system \eqref{hamiltonian}. Such a classical motion is, in turn, completely fixed by the the Floquet modes $\exp(iWt)$ and the Floquet-Lyapunov transformation $F(t)$.

The intelligent states \cite{trifonov1,TRIFO} of a quantum mechanical system can be  defined as those that minimize the Robertson uncertainty relation \eqref{intelligencecondition} for all times. Since this inequality is an extension of Heisenberg and Schr\"odinger inequalities, the intelligent states may be interpreted as a generalization of the canonical coherent states. In what follows we  will build the intelligent states of the N-dimensional non-stationary system \eqref{hamiltonian}. But, let us first show that in this Fock basis, \eqref{dynfock}, the Robertson's inequality \eqref{intelligencecondition} is not minimized. Using \eqref{aonfock} and \eqref{adagaonfock} to calculate the mean values present in \eqref{covarianceelements} for the integrals of motion and making use of the orthonormality of the Fock dynamical basis, we obtain 
\begin{equation}
	\sigma_{Fock}(A_i^\dagger,A_j)=\langle A_i^\dagger A_j\rangle-\langle A_i^\dagger\rangle\langle A_j\rangle+\dfrac{\delta_{ij}}{2}
\end{equation}
\begin{equation*}
	=\langle A_i^\dagger A_j^\dagger\rangle+\dfrac{\delta_{ij}}{2}=(n_{i}+\dfrac{1}{2})\delta_{ij}
\end{equation*}
The covariance matrix for $\hat{A}^\dagger(t)$ and $\hat{A}(t)$ in the Fock dynamical states is
\begin{equation*}
	\sigma_{n;t}(\hat{I}(t))=\left(\begin{array}{cc}
		0 & diag(\{n\})+1_N/2 \\
		diag(\{n\})+1_N/2 & 0
	\end{array}
	\right),
\end{equation*}
and for the canonical observables
\begin{equation}\label{fockcov}
	\sigma_{n;t}(\hat{\textbf{x}})=\frac{1}{2}F(t)F^\dagger(t)+Fdiag(\{n\})F^\dagger.
\end{equation}
Notice that the determinant of this matrix will increase with the number of excitations. This means that, as in the non-parametric problem, in general, these Fock states do not minimize the uncertainty relation and are non-classical states of the system.  

We define then, the coherent states of the system of interacting paramateric oscillators, through the action of the displacement operator $\mathcal{D}(\{\alpha\})$  on the ground state $|\{0\};t\rangle$ 
\begin{equation*}
	|\{\alpha\};t\rangle=\mathcal{D}(\{\alpha\})|\{0\};t\rangle=
\end{equation*}
\begin{equation}\label{cs}
	=\prod_{i=1}^N \exp\left(\alpha_i\hat{A}_i^\dagger(t)+\alpha_i^*\hat{A}_i(t)\right)|\{0\};t\rangle.
\end{equation}
From \eqref{cs} it follows that $|\{\alpha\};t\rangle$ are eigenstates of the linear integral of motion $\hat{A}(t)$,
\begin{equation}
	\hat{A}_i(t)|\{\alpha\};t\rangle=\alpha_{i}|\{\alpha\};t\rangle,
\end{equation}
where $\alpha_i$ is  equal to the initial value of the classical Floquet mode $\chi_{i+N}(0)$. This expression can be used to calculate the coordinate representation wave function \cite{landa,holz}.
It is easy to see that, as in the simple harmonic oscillator, the mean values of the positions $\hat{q}_i$ and momenta $\hat{p}_i$ on these states can reproduce the behavior of any solutions to the classical equation of motion. The mean values of the quadrature operators in coherent states are given by
\begin{equation*}
	\langle\hat{x}_i\rangle_{CS}=\sum_{j=1}^{2N}F_{ij}(t)\chi_j(t),
\end{equation*}
or, more explicitly,
\begin{equation}
	\langle\hat{q}_i\rangle_{CS}=\sum_{j=1}^{N}\left(U_{ij}(t)\chi_j(0)e^{i\omega_j t}+U_{ij}^{*}(t)\chi_{j+N}(0)e^{-i\omega_j t}\right),
\end{equation}
which is exactly the general Floquet solution to the classical equations of motion. In these expressions $\chi(t)$ denotes the classical Floquet modes, they have the same functional form as the operators \eqref{floquetmodes} of the Heisenberg picture.

We now find an explicit expression for the covariance matrices of the integrals of motion and canonical observables in the coherent states \eqref{cs}. To that end we begin by noticing that, since $\mathcal{D}^{\dagger}(\{\alpha\})=\mathcal{D}^{-1}(\{\alpha\})$ and
\begin{equation*}
	\mathcal{D}^{\dagger}(\{\alpha\})\hat{A}(t)\mathcal{D}(\{\alpha\})=\hat{A}(t)+\alpha,	
\end{equation*} 
\begin{equation*} 
\mathcal{D}^{\dagger}(\{\alpha\})\hat{A}^\dagger(t)\mathcal{D}(\{\alpha\})=\hat{A}^\dagger(t)+\alpha^*,
\end{equation*} 
the covariances between the integrals of motion in a given coherent state are invariant under the action of the displacement operator. This is a major simplification in our demonstration, it allows us to consider only the covariance matrix for the ground state of the system, which is shared by all coherent states. Then, for any values of $\{\alpha\}$, the covariance matrix will be given by setting $n_i=0,\hspace*{0.2cm}i=1,...,N$ in \eqref{fockcov}. Written explicitly, for the integrals of motion, it reads
\begin{equation}\label{covmatCSliom}
	\sigma_{\alpha;t}(\hat{I}(t))=\left(\begin{array}{cc}
		0 & 1_N/2 \\
		1_N/2 & 0
	\end{array}
	\right)
\end{equation}
and for canonical  observables,
\begin{equation}\label{covCS}
	\sigma_{\alpha;t}(\hat{\textbf{x}})=\frac{1}{2}F(t)F^\dagger(t)=\sigma_{0;t}(\hat{\textbf{x}})
\end{equation}
As in the Fock case \eqref{fockcov}, the Floquet exponents cancel out and $F(t)$ is sufficient to know the covariances between the position and momentum operators. This expression shows that the coherent states \eqref{cs}, of the non-stationary system \eqref{hamiltonian} are squeezed and correlated.

Squeezing occurs when the uncertainty in position is different from the uncertainty in momentum. Even in the one-dimensional case, the parametric excitation induces squeezing of the quadratures components \cite{manko}. This can be directly seen from \eqref{covCS}: the position uncertainty for the $i-$th oscillator ($i<N$) is given by the $i$-th main diagonal element of $FF^\dagger$. On the other hand, the momentum uncertainty corresponds to the $i+N$ element of the diagonal, since these two are in general different, so are the uncertainties. 

They are correlated in a double sense. Due to the coupling between the position and momentum operators in the Hamiltonian, it is impossible to express the coherent state $|\{\alpha\};t\rangle$ of the entire system, as a product of the single oscillators coherent states $|\alpha_i;t\rangle$. But also, because the parametric excitation guarantees finite correlations between position and momentum operators in each independent oscillator.

To finish this section we want to prove the intelligent character of the (generalized) coherent states \eqref{cs}. In other words, we have to demostrate that $\det\left(\sigma_{\alpha;t}(\hat{\textbf{x}})\right)=2^{-2N}$. This requires an expression for the determinant of the Floquet-Lyapunov transformation subject to the canonical condition \eqref{cannonicalcondition}. This is a rather simple task if one notices that the matrix 
\begin{equation}\label{symplectictransf}
	\tilde{F}(t)=\left(\begin{array}{cc}
		1_N & 0 \\
		0 & i1_N
	\end{array}
	\right)F^T(t),
\end{equation}
is an element of the complex symplectic group $Sp(\mathcal{C},2n)$, which guarantees that its determinant is equal one. Then
\begin{equation}
	det[F(t)]=det[\left(\begin{array}{cc}
		1_N & 0 \\
		0 & -i1_N
	\end{array}
	\right)]=(-i)^{N}.
\end{equation}
It follows directly from \eqref{covmatCSliom} that the determinant of the covariance matrix for the linear integrals of motion, in a coherent state, is equal to $(-4)^{-N}$. We then substitute last two results in \eqref{sigmaF} to obtain that the determinant of $\sigma(\hat{x})$ is indeed equal to $2^{-2N}$. This concludes the proof that  the coherent states $|\{\alpha\};t\rangle$ of the parametrically excited oscillators minimize the Robertson uncertainty relation \eqref{intelligencecondition} for any configuration $\Pi(t)$ of the system.

In summary, we  have related the (quantized) Floquet modes of the classical system with the linear integrals of motion \eqref{integralsofmotion} of the quantum mechanical problem (\ref{hamiltonian}), showing that the work done in \cite{landa} for a set of coupled Mathieu equations can be extended to any configuration of oscillators with time-periodic parameters. The Floquet-Lyapunov transformation can be seen as a multidimensional and time-dependent generalization of the usual relations between the creation and annihilation operators (corresponding to Floquet modes) and the position and momentum operators of a simple harmonic oscillator. Moreover, a proper use of the displacement operator allows the derivation of the coherent states of the model. We also showed that the explicit time-dependence of covariances between canonical operators, in coherent and Fock states, is completely determined by this periodic transformation. If we were to consider a stationary system described by $\Pi(t_0)$ in \eqref{hamiltonian}, for a fixed $t_0$, the intelligent states \eqref{cs} reduce to the multimode minimum uncertainty squeezed states derived in \cite{Milburn}. In the next section, we will show that this framework can be used to construct a special case of non-Gaussian states.

\section{Adding quanta to intelligent states}

The action of the creation operator, or photon addition, on the ground sate of a harmonic oscillator generates a set of non-classical states, the Fock basis. In analogy, the photon-added or excited coherent states (PACS) are defined as the normalized states obtained through the addition of photons to the canonical coherent states. In the nonstationary setting \eqref{hamiltonian} of coupled oscillators, the role of the creation operator is played by the integral of motion $\hat{A}^\dagger(t)$, and the dynamical version of the PACS is constructed through its action on the intelligent states \eqref{cs}, 
\begin{equation}\label{PACS}
	|\{\alpha\},\{m\};t\rangle=\prod_{k=1}^N\dfrac{(\hat{A}_k^\dagger(t))^{m_k}}{\sqrt{m_k!L_{m_k}(-|\alpha_k|^2)}}|\{\alpha\};t\rangle
\end{equation}
where $L_{m_k}(-|\alpha_k|^2)$ is the Laguerre polynomial of order $m_k$. If $m_k=0$, for all $k$, we recover the coherent states $|\{\alpha\},\{0\};t\rangle=|\{\alpha\};t\rangle$, and Fock states are recovered by setting $\alpha_k=0$, for all $k$, since $L_{m_k}(0)=1$. 

The wave function describing these states can be directly computed from their definition (see supplementary material). The result is  
\begin{equation}\label{PACSwf}
	\psi_{\alpha,m}(\textbf{q},t)=N_{\alpha,m}\psi_{\alpha}(\textbf{q},t)\mathcal{H}_{\textbf{m}}^{U^{\dagger}U^{-T}}\left(U^{-*}\textbf{q}-\vec{\alpha}\right).
\end{equation}
In this expression $N_{\alpha,m}$ is the normalization constant appearing in \eqref{PACS}, $\psi_{\alpha}$ denotes the wavefunction of the intelligent states \cite{holz,landa} and $\mathcal{H}_{\textbf{m}}$ is the multidimensional Hermite polynomial \cite{bateman} of order $\textbf{m}=(m_1,...m_N)$, defined by the relation
\begin{equation}
	\mathcal{H}_{\textbf{m}}^{M}\left(\textbf{q}\right)=(-1)^{\Sigma m_k}e^{\frac{1}{2}\textbf{q}^TM\textbf{q}}\dfrac{\partial^{\Sigma m_k}}{\partial q_1...\partial q_N}e^{-\frac{1}{2}\textbf{q}^TM\textbf{q}}
\end{equation} 
Expression \eqref{PACSwf} is a generalization of single oscillator wavefunction obtained in \cite{dodonovPacs} for the PACS. Due to the mode coupling, the wavefunction is not expressible as a product of single oscillators wavefunctions. This can be directly seen from the definition of the multidimensional Hermite polynomials, which cannot be expressed as the product of unidimensional polynomials if the matrix $U^{\dagger}U^{-T}$ is not diagonal. 

From equation \eqref{PACSwf} we can obtain the Wigner function of the photon-added states \eqref{PACS}. The derivation is sketched in the suplementary material. The result is
\begin{equation*}
	W_{PACS}(\textbf{q},\textbf{p},t)=2^N\exp\left(- 2 |A(\textbf{q},\textbf{p},t)-\alpha|^2\right)\times
\end{equation*}
\begin{equation}\label{Wignerfunction}
	\times\left(\prod_{k=1}^N(-1)^{m_k}\dfrac{L_{m_k}(|2A(\textbf{q},\textbf{p},t)-\alpha|_k^2)}{L_{m_k}(-|\alpha_k|^2)}\right)
\end{equation}
The \emph{function} $A(\textbf{q},\textbf{p},t)$ has the same form \eqref{liomintermsofqp} as the integral of motion $\hat{A}(t)$,
\begin{equation}
		A(\textbf{q},\textbf{p},t)=-iV^T(t)e^{i\Omega t}\textbf{q}+iU^T(t)e^{i\Omega t}\textbf{p}.
\end{equation}
If we set $m_k=0$ for every value of $k$, then \eqref{Wignerfunction} reduces to a Gaussian distribution. Note that, when coupling is present, \eqref{Wignerfunction} cannot be written as the product of single mode Wigner functions. 

This fact has an interesting consecuence that does not arise in single mode cases \cite{agarwal,dodonovPacs}. To illustrate this we can consider two coupled oscillators ($N=2$) in an intelligent state \eqref{cs}, described by the eigenvalues $\alpha_1$ and $\alpha_2$. The Wigner function is given by \eqref{Wignerfunction}, with $m_1=m_2=0$, and it takes no negative values for any combination of $\alpha_1$ and $\alpha_2$, such state can then be considered a classical state. If we add a single excitation ($m_2=1$) to only one of them, the Wigner function will become
\begin{equation}
	4\frac{|2A_1(q_1,p_1,t)-\alpha_1|^2-1}{1+|\alpha_1|^2}e^{-2|A(\textbf{q},\textbf{p},t)-\alpha|^2},
\end{equation}
and it can take negative values whenever $|2A_1(q_1,p_1,t)-\alpha_1|^2<1$. The mode $\alpha_1$, to which we added no excitations ($m_1=0$), cannot be said to be in a classical state, since its corresponding canonical observables are correlated to those of the mode that was excited. This is caused by the coupling between the oscillators and its true for any number of excitations.

To calculate the covariances we employ the following expression for the expected value of a generic operator $\hat{o}$ on these states
\begin{equation*}
	\langle\hat{o}\rangle_{\{\alpha\},\{m\};t}=\dfrac{e^{-|\vec{\alpha}|^2}}{\prod m_k!L_{m_k}}\prod_{i=1}^N\sum_{\{n_i,\bar{n}_i\}=0}^{\infty}\dfrac{(\alpha_i)^{n_i}(\alpha_i^*)^{\bar{n}_i}}{\sqrt{n_i!\bar{n}_i!}}\times
\end{equation*}
\begin{equation*}
	\times\left(\sqrt{n_i+1}...\sqrt{n_i+m_i}\right)\left(\sqrt{\bar{n}_i+1}...\sqrt{\bar{n}_i+m_i}\right)\times
\end{equation*}
\begin{equation}\label{omeanvalue}
	\times\langle n_1+m_1,...;t|\hat{o}|\bar{n}_1+m_1,...;t\rangle.
\end{equation}
This means that to obtain the covariances between the integrals of motion in photon-added coherent states, we first need to calculate the mean values of such integrals and the matrix elements of the products between them in the dynamical Fock basis. For the sake of simplicity we will define
\begin{equation*}
	\langle n_1+m_1,...;t|\hat{o}|\bar{n}_1+m_1,...;t\rangle=\langle...\rangle_{Fock}
\end{equation*}
and we will also omit the explicit time-dependence. The needed expressions are readily obtained from \eqref{aonfock} and \eqref{adagaonfock},
\begin{equation}
	\langle\hat{A}_i\rangle_{Fock}=\sqrt{n_i+m_i}\delta_{n_i,n_i-1}
\end{equation} 
\begin{equation}
	\langle\hat{A}^\dagger_i\rangle_{Fock}=\sqrt{n_i+m_i+1}\delta_{n_i,n_i+1}
\end{equation} 
\begin{equation}
	\langle\hat{A}_i\hat{A}_j\rangle_{Fock}=\left\{\begin{array}{l}
		\langle\hat{A}_i\rangle\langle\hat{A}_j\rangle\hspace*{0.3cm}i\neq j \\
		\sqrt{n_i+m_i}\sqrt{n_i+m_i-1}\times \\
		\times\delta_{n_i,n_i-2},\hspace*{0.3cm}i=j
	\end{array}\right.
\end{equation}
\begin{equation}
	\langle\hat{A}^\dagger_i\hat{A}^\dagger_j\rangle_{Fock}=\left\{\begin{array}{l}
		\langle\hat{A}^\dagger_i\rangle\langle\hat{A}^\dagger_j\rangle\hspace*{0.3cm}i\neq j \\
		\sqrt{n_i+m_i+1}\sqrt{n_i+m_i+2}\times \\
		\times\delta_{n_i,n_i+2},\hspace*{0.3cm}i=j
	\end{array}\right.
\end{equation}
\begin{equation}
	\langle\hat{A}^\dagger_i\hat{A}_j\rangle_{Fock}=\left\{\begin{array}{l}
		\langle\hat{A}^\dagger_i\rangle\langle\hat{A}_j\rangle\hspace*{0.3cm}i\neq j \\
		(n_i+m_i),\hspace*{0.3cm}i=j
	\end{array}
	\right.
\end{equation}
The corresponding expression for $\langle\hat{A}_i\hat{A}^\dagger_j\rangle$ is obtained from the commutation relations \eqref{weylhei}, $\langle\hat{A}_i\hat{A}^\dagger_j\rangle=\delta_{ij}-\langle\hat{A}_j\hat{A}^\dagger_i\rangle$. We then substitute these results in \eqref{omeanvalue} to compute the analogous expressions in the photon-added coherent states. For clarity, let us calculate the mean value of the integral of motion $\hat{A}$,
\begin{equation*}
	\langle\hat{A}_i\rangle_{PACS}=\dfrac{e^{-|\alpha_i|^2}(m_i!)^{-1}}{L_{m_i}(-|\alpha_i|^2)}\sum_{n=0}^{\infty}\dfrac{|\alpha_i|^{2n}}{n!}(n+1)...(n+m_i+1).
\end{equation*}



Introducing the \emph{Pochhammer symbol}: $	(a)_b=1\cdot2\cdot3...(a+b-1),\hspace*{0.3cm}a,b\in Naturals,$
it can be written as
\begin{equation}
	\langle\hat{A}_i\rangle_{PACS}=\alpha_i\dfrac{e^{-|\alpha_i|^2}}{L_{m_i}(-|\alpha_i|^2)}\sum_{n=0}^{\infty}\dfrac{|\alpha_i|^{2n}}{n!}\dfrac{(m_i+2)_{n}}{(2)_{n}}.
\end{equation}
Identifying the sum on the right hand side with the confluent hypergeometric function $_{1}F_{1}(m_i+2,2;-|\alpha_i|^2)$, 
\begin{equation}
	\langle\hat{A}_i\rangle_{PACS}=\alpha_i\dfrac{e^{-|\alpha_i|^2} {_{1}}F_{1}(m_i+2,2;-|\alpha_i|^2)}{L_{m_i}(-|\alpha_i|^2)}
\end{equation}
We now make use of Kummer's transformation, $	_{1}F_{1}(a,c;x)=e^{x}{_{1}}F_1(c-a,c;-x),$ and the relation $_1F_1(-n,a+1;x)=\dfrac{n!}{(a+1)_n}L_{n}^a(x)$
to finally obtain
\begin{equation}\label{aniPACS}
	\langle\hat{A}_i\rangle_{PACS}=\alpha_i\dfrac{L_{m_i}^1(-|\alpha_i|^2)}{L_{m_i}(-|\alpha_i|^2)}
\end{equation} 
Proceeding similar for the creation operators:
\begin{equation}\label{crePACS}
	\langle\hat{A}^\dagger_i\rangle_{PACS}=\alpha_i^*\dfrac{L_{m_i}^1(-|\alpha_i|^2)}{L_{m_i}(-|\alpha_i|^2)}
\end{equation} 
We can now use \eqref{aniPACS} and \eqref{crePACS}, together with \eqref{FLT} to calculate the mean value of the quadrature operators
\begin{equation*}
	\langle\hat{x}_i\rangle_{PACS}=\sum_{j=1}^{2N}F_{ij}(t)\chi_j(t)\dfrac{L_{m_j}^1(-|\alpha_j|^2)}{L_{m_j}(-|\alpha_j|^2)}
\end{equation*}
where $\chi_j(t)$ is the $j$th (classical) Floquet mode. If $m=0$ we recover the mean values in the coherent states. In practice, the effect of this photon-addition on the mean values of the quadrature components is a "deformation" of the classical phase space trajectory of the coherent states. We could see this as a change of the initial conditions of the classical motion, which is reflected in the constant initial values of $\chi$ as 
\begin{equation}
	\left\{\chi_j(0)\right\}_{PACS}=\left\{\chi_j(0)\right\}_{CS}\dfrac{L_{m_j}^1(-|\alpha_j|^2)}{L_{m_j}(-|\alpha_j|^2)}
\end{equation}


The second order statistical moments of the integrals of motion are calculated in a similar way:
\begin{equation}
	\langle\hat{A}_i\hat{A}_j\rangle_{PACS}=\left\{\begin{array}{l}
		\langle\hat{A}_i\rangle\langle\hat{A}_j\rangle\hspace*{0.3cm}i\neq j \\
		(\alpha_i)^2\dfrac{L_{m_i}^2(-|\alpha_i|^2)}{L_{m_i}(-|\alpha_i|^2)}\hspace*{0.3cm}i=j
	\end{array}\right.
\end{equation}
\begin{equation}
	\langle\hat{A}^\dagger_i\hat{A}^\dagger_j\rangle_{PACS}=\left\{\begin{array}{l}
		\langle\hat{A}_i^\dagger\rangle\langle\hat{A}_j^\dagger\rangle\hspace*{0.3cm}i\neq j \\
		(\alpha_i^*)^2\dfrac{L_{m_i}^2(-|\alpha_i|^2)}{L_{m_i}(-|\alpha_i|^2)}\hspace*{0.3cm}i=j
	\end{array}\right.
\end{equation}
\begin{equation}
	\langle\hat{A}^\dagger_i\hat{A}_j\rangle_{PACS}=\left\{\begin{array}{l}
		\langle\hat{A}_i^\dagger\rangle\langle\hat{A}_j\rangle\hspace*{0.3cm} i\neq j \\
		|\alpha_i|^2\dfrac{L_{m_i}^1(-|\alpha_i|^2)}{L_{m_i}(-|\alpha_i|^2)}+m_l \hspace*{0.3cm}i=j
	\end{array}
	\right.
\end{equation}


From these relations follow that all non-diagonal elements of each block of the covariance matrix (for the creation and annihilation operators) are zero, and we only need to calculate the diagonal ones, using \eqref{covarianceelements}. To avoid a cumbersome notation we define
\begin{equation}\label{p}
	P_{m_i}^{n}(-|\alpha_i|^2)\equiv\dfrac{L_{m_i}^{n}(-|\alpha_i|^2)}{L_{m_i}(-|\alpha_i|^2)}-\left(\dfrac{L_{m_i}^{1}(-|\alpha_i|^2)}{L_{m_i}(-|\alpha_i|^2)}\right)^2
\end{equation}
The elements of the covariance matrix for the integrals of motion $\hat{I}(t)$ in $|\{\alpha\},\{m\};t\rangle$ are then given by
\begin{equation*}
	\sigma_{PACS}(A_i^\dagger(t),A_i^\dagger(t))=(\alpha_i^*)^2P_{m_i}^2(-|\alpha_i|^2),
\end{equation*}
\begin{equation*}
	\sigma_{PACS}(A_i,A_i)=(\alpha_i)^2P_{m_i}^2(-|\alpha_i|^2),
\end{equation*}
\begin{equation}\label{covPACS}
	\sigma_{PACS}(A_i^\dagger,A_i)=|\alpha_i|^2P_{m_i}^1(-|\alpha_i|^2)+m_i+1/2,  
\end{equation}
The covariance matrix formed with \eqref{covPACS} reduces to that of Fock states \eqref{fockcov} if one takes $\alpha_k=0$, and to that of coherent states if $m_k=0$, for all values of $k$, as it should be. 
The explicit appearance of $\alpha_i$ and $m_i$ in \eqref{covPACS} implies that the covariances depend on both the coherent state that was excited and the number of excitations added to it. Although all coherent states $|\{\alpha\};t\rangle$ share the same covariance matrix \eqref{covCS}, the states obtained by their excitation $|\{\alpha\},\{m\};t\rangle$ do not. Since the excitation of the ground state ($\alpha=0$) produces Fock states, which are non-classical and do not minimize the uncertainty relation for $n\neq0$, also the excitation of the coherent states produces states that are both non-classical and non-intelligent for any combination of $\{\alpha\}$ and $\{m\}$ ($m\neq0$).  

The corresponding covariance matrix for the position and momentum operators $\sigma_{PACS}(\hat{x})$ is given by direct substitution of \eqref{covPACS} in \eqref{sigmaF}. After some algebraic manipulations and using \eqref{floquetmodes} and \eqref{ffconjugate}, we find 
\begin{equation*}
	\sigma_{PACS}(x_i,x_j)=\sum_{k=1}^{2N}F_{ik}F_{jk}\chi_k^2(t)P_{m_k}^{(2)}(-|\alpha_k|^2)+
\end{equation*}
\begin{equation*}
	+\sum_{k=1}^N\left\{(F_{ik}F_{j,k+N}+F_{i,k+N}F_{j,k})\times\right.
	\end{equation*}
\begin{equation}
	\left.\times(|\alpha_i|^2P_{m_i}^1(-|\alpha_i|^2)+m_i+1/2)\right\}.
\end{equation}
The functions $\chi_k(t)$ appearing in the first term of the right hand side, are the classical Floquet modes $\chi(0)\exp(iWt)$. The initial values $\chi_k(0)$ are equal to $\alpha_k^*$, for $k<N$ and to $\alpha_k$ for $k>N$. This is a major difference between the covariances in PACS and the covariances in intelligent states \eqref{covCS}, or Fock states \eqref{fockcov}. In these two last, the covariances between canonical operators are totally determined by the Floquet-Lyapunov transformation and, in the case of Fock states, by the number of excitations. In the photon-added states, on the other hand, to fully determine the covariances we also need the mean number of photons $|\alpha_k|^2$ (of the intelligent states) and the Floquet modes of the classical system.

It is important to note that even if the covariances between position and momentum operators in a photon-added coherent state are functions of time, the \emph{determinant} of the covariance matrix is $constant$, it is not affected by the parametric excitation. This implies that the difference between such a determinant and the minimum of the Robertson relation is only a function of $\alpha_i$ and $m_i$ and does not depend on the Floquet-Lyapunov transformation. This result is a multidimensional generalization of that obtained in \cite{dodonovPacs} for the uni-dimensional system.

\section{Conclusions}

In this work we have determined the photon-added coherent states of a system of coupled oscillators whose frequencies and coupling factors change periodically with time. Exploiting the Floquet-Lyapunov theory we derived first, the linear  integrals of motion of this system and related them to the quantum mechanical counterpart of the Floquet modes in the classical theory. The Floquet-Lyapunov transformation was shown to provide a link between the covariance matrices of the integrals of motion and of canonical observables through expression \eqref{sigmaF}. In other words, the time evolution of the covariance matrix for the canonical observables is fully determined by that of the Floquet-Lyapunov  transformation and the Floquet modes of the classical system.

Furthermore, we showed that the integrals of motion of the problem can be easily translated into operators that behave as the creation and annihilation operators of the harmonic oscillator and we used this analogy to obtain the (generalized) coherent states of this system. These were shown to be the intelligent states, that is, they minimize the Robertson's inequality. Finally, this analogy was further extended to construct the photon-added coherent states and to obtain their wavefunction in coordinate representation and the corresponding Wigner distribution. These are not-intelligent states, for which the determinant of the covariance matrix depends on both the complex eigenvalues of the annihilation operator that describes the coherent state and the number of excitations added to it. We obtained that, although the elements of the covariance matrix for canonical observables are explicitly time-dependent, the determinant of the covariance matrix is a constant strictly greater than the Robertson's minimum, for nonzero $m$. 

We want to emphasize that these results are valid for any periodic quadratic Hamiltonian of the form \eqref{hamiltonian}, as long as the corresponding classical system is stable. Moreover, we are confident that our findings might be of interest to attack continuous-variable quantum information problems, where analytical results concerning non-Gaussian states are not as common as for the Gaussian counterpart. For example, the wavefunction \eqref{PACSwf} can be used to construct phase space representations of the PACS, photon distribution functions and in general any quantity of importance for quantum optical applications in time-periodic settings.

\acknowledgments
We want to thank E. Aurell, A. Buchleitner and M. Walschaers for the careful reading of the manuscript and useful suggestions.

\section{Suplementary Material}
We show here the detailed calculation of the photon-added states wavefunction in coordinate representation. That is, we find
\begin{equation*}
	\psi_{\alpha,m}(\textbf{q},t)=\langle\textbf{q}|\{\alpha\},\{m\};t\rangle
\end{equation*}
From \eqref{PACS} it follows that
\begin{equation*}
	\psi_{\alpha,m}(\textbf{q},t)=N_{m}e^{-|\alpha|^2/2}\times
\end{equation*}
\begin{equation*}
	\times\sum_{n}\prod_{i=1}^N\dfrac{(\alpha_i)^{n_i}}{\sqrt{n_i!}}\sqrt{(n_i+1)...(n_i+m_i)}\langle\textbf{q}|\textbf{n}+\textbf{m}\rangle
\end{equation*}
\begin{equation*}
	=N_{n,m}e^{-|\alpha|^2/2}e^{\frac{1}{2}i\textbf{q}^TU^{-T}V^T\textbf{q}}\sum_{n}\prod_{i=1}^N\dfrac{(\alpha_i)^{n_i}}{\sqrt{n_i!}}\mathcal{H}_{\textbf{m}+\textbf{n}}^M(U^{-*}\textbf{q})
\end{equation*}
We have to calculate the sum on the right hand side of this expression. Using the definition of the multidimensional Hermite polynomials, the sum can be recast as
\begin{equation}
	(-1)^{\Sigma_i m_i}e^{\frac{1}{2}\textbf{q}^TU^{-\dagger}MU^{-*}\textbf{q}}\dfrac{\partial^{\Sigma_i m_i}}{\partial q_{1}^{m_1}...\partial q_{N}^{m_N}}\times
\end{equation}
\begin{equation*}
	\times\left(e^{-\frac{1}{2}\textbf{q}^TU^{-\dagger}MU^{-*}\textbf{q}}\sum_{\{n\}}\prod_{i=1}^N\dfrac{(\alpha_i)^{n_i}}{\sqrt{n_i!}}\mathcal{H}_{\textbf{n}}^M(U^{-*}\textbf{q})\right)
\end{equation*}
The term between parenthesis contains the generating function for Hermite polynomials, so it will transform into
\begin{equation*}
	e^{-\frac{1}{2}\textbf{q}^TU^{-\dagger}MU^{-*}\textbf{q}+\alpha^TMU^{-*}\textbf{q}-\frac{1}{2}\alpha^TM\alpha}
\end{equation*}
Completing squares on the exponent we obtain for the sum
\begin{equation*}
	e^{-\frac{1}{2}\alpha^TM\alpha+\alpha^TMU^{-*}\textbf{q}}\mathcal{H}_{\textbf{n}}^M(U^{-*}\textbf{q}-\alpha)
\end{equation*}
We substitute this in the expression for $\psi_{\alpha,m}$ and recognize
\begin{equation*}
	\psi_{\alpha}(\textbf{q},t)=N_{\alpha}e^{-|\alpha|^2/2}e^{\frac{1}{2}i\textbf{q}^TU^{-T}V^T\textbf{q}-\frac{1}{2}\alpha^TM\alpha+\alpha^TMU^{-*}\textbf{q}}
\end{equation*}
as the wavefunction of the intelligent states \cite{landa,holz}. This will lead us to the result \eqref{PACSwf}.
The Wigner function is constructed from
\begin{equation*}
	W(\textbf{q},\textbf{p},t)=\int_{R^{N}}d^{N}\textbf{r}e^{-i\textbf{p}^T\textbf{r}}\psi_{\alpha,m}(\textbf{q}+\frac{\textbf{r}}{2},t)\psi_{\alpha,m}^*(\textbf{q}-\frac{\textbf{r}}{2},t)
\end{equation*}
This will transform into an integral of Gaussian weight function times the product of two N-dimensional Hermite polynomials. To solve it, we can write such a product as
\begin{equation*}
	\mathcal{H}_{\textbf{m}}^{M}\left(\textbf{v}\right)\mathcal{H}_{\textbf{m}}^{M^*}\left(\textbf{w}^*\right)=\mathcal{H}_{\textbf{m},\textbf{m}}^{Q}\left(\textbf{z}\right)
\end{equation*}
where 
\begin{equation*}
	Q=\left(\begin{array}{cc}
		M & 0 \\
		0 & M^*
	\end{array}\right)	
\end{equation*}
and $\textbf{z}^T=(\textbf{v}^T\hspace*{0.2cm}\textbf{w}^\dagger)$. We then introduce 2N arbitrary parameters and make use of the generating function of multidimensional Hermite polinomials to express the integral as a Gaussian integral. The result will contain a product of N two-dimensional Hermite polynomials whos defining matrix is the Pauli matrix 
\begin{equation*}
	\left(\begin{array}{cc}
		0 & 1 \\
		1 & 0
	\end{array}
	\right).
\end{equation*}
This will allow us to express them as standard Laguerre polynomials of order $m_k$, and we will obtain \eqref{Wignerfunction}.

\bibliographystyle{eplbib}
\bibliography{Mybib3.bib}

\begin{thebibliography}{10}
\expandafter\ifx\csname url\endcsname\relax\def\url#1{\texttt{#1}}\fi

\bibitem{ncoupldho}
\Name{Urz\'ua A.~R., Ramos-Prieto I., Fern\'andez-Guasti M. \and Moya-Cessa
  H.~M.} \REVIEW{Quantum Reports}{1}{2019}{}.

\bibitem{tdhopointtrans}
\Name{Zelaya K. \and Rosas-Ortiz O.} \REVIEW{arXiv:1909.01948}{}{2019}{}.

\bibitem{landa}
\Name{Landa H., Drewsen M., Reznik B. \and Retzker A.} \REVIEW{Journal of
  Physics A-Mathematical and Theoretical}{45}{2012}{}.

\bibitem{wolf1}
\Name{Wolf K.~B.} \REVIEW{J. Appl. Math.}{40}{1981}{}.

\bibitem{tdhocs}
\Name{Hartley J.~G. \and Ray J.~R.} \REVIEW{Phys. Rev. D}{25}{1982}{}.

\bibitem{holz}
\Name{Holz A.} \REVIEW{Lettere Al Nuovo Cimento}{4}{1970}{}.

\bibitem{manko}
\Name{Dodonov V.~V., Man'ko O.~V. \and Man'ko V.~I.} \REVIEW{Proceedings of the
  Lebedev Physical Institute}{191}{1989}{}.

\bibitem{variablemedia}
\Name{Kryuchkov S.~I., Suazo E. \and Suslov S.~K.}
  \REVIEW{arXiv:1401.2924}{}{2014}{}.

\bibitem{kashmir}
\Name{Dodonov V.~V. \and Dodonov A.~V.} \REVIEW{Journal of Russian Laser
  Research}{26}{2005}{}.

\bibitem{rmpblatt}
\Name{Leibfried D., Blatt R., Monroe C. \and Wineland D.} \REVIEW{Reviews of
  Modern Physics}{75}{2003}{}.

\bibitem{quantumcomputation}
\Name{Cirac J.~I. \and Zoller P.} \REVIEW{Phys. Rev. Lett.}{74}{1995}{}.

\bibitem{kalmykov}
\Name{Kalmykov S.~Y. \and Veisman M.~E.} \REVIEW{Phys. Rev. A}{57}{1998}{}.

\bibitem{lizuain}
\Name{Lizuain I., Palmero M. \and Muga J.~G.} \REVIEW{Phys. Rev.
  A}{95}{1996}{}.

\bibitem{glauber}
\Name{Glauber R.~J.} \REVIEW{Phys. Rev. Lett.}{10}{1963}{}.

\bibitem{heisenberg}
\Name{Heisenberg W.} \REVIEW{Zeitschrift für Physik}{43}{1927}{}.

\bibitem{kennard}
\Name{Kennard E.~H.} \REVIEW{Z. für Phys.}{44}{1927}{}.

\bibitem{kalmykov2}
\Name{Veisman M.~E. \and Kalmykov S.~Y.} \REVIEW{Theoretical and Mathematical
  Physics}{111}{1997}{}.

\bibitem{schrodinger}
\Name{Schrödinger E.} \REVIEW{Sitzungberichten der Preussischen Akademie der
  Wissenschaften}{19}{1932}{}.

\bibitem{robertson}
\Name{Robertson H.~P.} \REVIEW{Phys. Rev.}{46}{1934}{}.

\bibitem{continousvariableqc}
\Name{Adesso G., Ragy S. \and Lee A.~R.} \REVIEW{arXiv:1401.4679}{}{2014}{}.

\bibitem{mandelparameter}
\Name{V~Buzek, A Vidiella-Barranco P. L.~K.} \REVIEW{Phys. Rev. A}{45}{1992}{}.

\bibitem{experimentalPACS}
\Name{Zavatta A., Viciani S. \and Bellini M.} \REVIEW{Science}{306}{2004}{}.

\bibitem{agarwal}
\Name{Agarwal G.~S. \and Tara K.} \REVIEW{Phys. Rev. A}{43}{1991}{}.

\bibitem{dodonovPacs}
\Name{Dodonov V.~V., Marchiolli M.~A., Korennoy Y.~A., Manko V.~I. \and Moukhin
  Y.~A.} \REVIEW{Phys. Rev. A}{58}{1998}{}.

\bibitem{MATTIA}
\Name{Walshaers M., Fabre C., Parigi V. \and Treps N.} \REVIEW{Phys. Rev.
  A}{96}{2017}{}.

\bibitem{gfloquet}
\Name{Floquet G.} \REVIEW{Annales scientifiques de l'École Normale
  Supérieure}{8}{1879}{}.

\bibitem{maclahan}
\Name{McLachlan N.~W.} \Book{Theory and Application of Mathieu Functions}
  (Dover) 1965.

\bibitem{wolf2}
\Name{Wolf K.~B.} \REVIEW{J. Math. Phys.}{15}{1974}{}.

\bibitem{dodonovLIOM}
\Name{Dodonov V.~V., Malkin I.~A. \and Man’ko V.~I.} \REVIEW{Int. J. Theor.
  Phys.}{14}{1975}{}.

\bibitem{trifonov1}
\Name{Trifonov D.~A.} \REVIEW{Journal of Physics A}{}{1997}{}.

\bibitem{TRIFO}
\Name{Trifonov D.~A.} \REVIEW{Proceedings of the IInd Int. Conference on
  Geometry, Integrability and Quantization}{}{2000}{}.

\bibitem{Milburn}
\Name{Milburn G.~J.} \REVIEW{J. Phys. A: Math. Gen.}{17}{1984}{}.

\bibitem{bateman}
\Name{Bateman H.} \Book{Higher Trascendental Functions} Vol.~2 (McGraw-Hill).

\end{thebibliography}
%
%

\end{document}